\documentclass[12pt,a4paper]{article}
\usepackage[cp866]{inputenc}
\usepackage{amsmath}
\usepackage{axodraw}
\usepackage[dvips]{epsfig}
\usepackage{graphicx}
\usepackage{epsfig}
\newcommand{\be}{\begin{equation}}
\newcommand{\ee}{\end{equation}}
\newcommand{\ben}{\begin{equation*}}
\newcommand{\een}{\end{equation*}}
\newcommand{\bea}{\begin{eqnarray}}
\newcommand{\eea}{\end{eqnarray}}
\newcommand{\ar}{\begin{array}}
\newcommand{\arn}{\end{array}}

\newcommand{\q}{\vec{q}}
\newcommand{\qs}{\vec{q}^{\;2}}

\def\pnot{\mbox{${\not{\hbox{\kern-3.0pt$p$}}}$}}
\def\qnot{\mbox{${\not{\hbox{\kern-2.0pt$q$}}}$}}
\def\enot{\mbox{${\not{\hbox{\kern-2.0pt$e$}}}$}}
\def\knot{\mbox{${\not{\hbox{\kern-2.0pt$k$}}}$}}

\def\fun#1#2{\lower3.6pt\vbox{\baselineskip0pt\lineskip.9pt\ialign
{$\mathsurround=0pt#1\hfil##\hfil$\crcr#2\crcr\sim\crcr}}}

\begin{document}
\sloppy                              
\renewcommand{\baselinestretch}{1.0} 

\begin{titlepage}
\hskip 11cm \vbox{ \hbox{Budker INP 2015-24}  }
\vskip 3cm

\begin{center}
{\bf On Hermitian separability of the  next-to-leading order
BFKL kernel  for the  adjoint representation of the gauge group  in the planar N = 4  SYM $^{\ast}$}
\end{center}

\centerline{V.S.~Fadin
$^{a,b\,\dag}$, R.~Fiore$^{c\,\ddag}$}

\vskip .6cm

\centerline{\sl $^{a}$
Budker Institute of Nuclear Physics of SD RAS, 630090 Novosibirsk
Russia}
\centerline{\sl\sl $^{b}$ Novosibirsk State University, 630090 Novosibirsk, Russia}
\centerline{\sl $^{c}$ Dipartimento di Fisica, Universit\`a della Calabria,}
\centerline{\sl and Istituto Nazionale di Fisica Nucleare, Gruppo collegato di Cosenza,}
\centerline{\sl Arcavacata di Rende, I-87036 Cosenza, Italy}

\vskip 2cm

\begin{abstract}
We analyze a modification of the  BFKL kernel for the  adjoint representation of the colour group in the maximally  supersymmetric (N=4) Yang-Mills theory   in the limit of a large number of colours, related to  the  modification of the eigenvalues of the kernel  suggested by S. Bondarenko and A. Prygarin in order to reach the Hermitian separability of the eigenvalues.  We restore the modified kernel in the momentum space. It turns out that the modification is related only to the real part of the kernel and that the correction to the kernel can not be presented by a single analytic function in the   entire momentum region, which contradicts the known properties of the kernel.     
\end{abstract}


\vfill \hrule \vskip.3cm \noindent $^{\ast}${\it Work supported 
in part by the Ministry of Education and Science of Russian Federation,
in part by  RFBR,  grant 16-02-00888, and in part by Ministero Italiano dell'Istruzione,
dell'Universit\`a e della Ricerca.} 
\vfill $
\begin{array}{ll} ^{\dag}\mbox{{\it e-mail address:}} &
\mbox{fadin@inp.nsk.su}\\
^{\ddag}\mbox{{\it e-mail address:}} &
\mbox{roberto.fiore@cs.infn.it}\\
\end{array}
$

\end{titlepage}

\section{Introduction}
The kernel of the BFKL (Balitsky-Fadin-Kuraev-Lipatov) equation   \cite{Fadin:1975cb}-\cite{Balitsky:1978ic} contains the so called real and virtual parts. The virtual part is determined by the gluon Regge trajectory and is the same for all representations of the colour group in the $t$-channel.  In the next-to-leading order (NLO)  the calculation of the trajectory  in QCD was carried out in  Refs.~\cite{Fadin:1995dd}-\cite{Fadin:1996tb} and was confirmed  in Refs.~\cite{Blumlein:1998ib,DelDuca:2001gu}. The supersymmetric Yang-Mills theories contain, in addition to the gauge bosons and fermions,  also scalar particles. Their contribution to the trajectory was obtained in Refs.~\cite{Kotikov:2000pm,Fadin:2007xy}.   The real part of the kernel comes from the real particle production. In QCD at the NLO  these particles are gluons and quark-antiquark pairs. Their contributions to the kernel for the adjoint representation of the colour group were calculated in Refs.~\cite{Fadin:2000kx,Fadin:2000hu} and  Ref.~\cite{Fadin:1998jv}  respectively.    The scalar particle contribution to the real part of the kernel was obtained in  Refs.~\cite{Gerasimov:2010zzb, Fadin:2007xy}. 

It is necessary to note here that the NLO corrections to the BFKL kernel are scheme dependent because of  the possibility to redistribute corrections to scattering amplitudes between the kernel and impact factors  of scattered particles \cite{Fadin:2006ha}. The calculations in Refs.~\cite{Fadin:2000kx}-\cite{Gerasimov:2010zzb}  were performed in the  scheme introduced in Ref.~\cite{Fadin:1998fv}, which we call the standard one.  It turns out, however, that in the N=4 supersymmetric Yang-Mills theory (N=4 SYM) in the planar limit  another scheme, which we call conformal, is more convenient. It is associated with the modified kernel ${\cal K}_m$, introduced in Ref.~\cite{Bartels:2008sc}, which is obtained from the usual BFKL kernel in the adjoint representation by subtraction of the gluon trajectory depending on the total $t$-channel momentum.  One of advantages of this  kernel is its infrared safety, which permits to consider this kernel at  physical transverse dimension $D-2 =2$. This advantage is manifested in all Yang-Mills theories.   Another  important  advantage,   manifested  in  the N=4 SYM, is the  dual conformal invariance, i.e. invariance under M\"{o}bius  transformations in the space of  dual two-dimensional transverse momenta.  In the leading order (LO) the invariance of $ {\cal K}_m$ is easily seen \cite{Bartels:2008sc}.  However, in the NLO in the standard scheme, in which the kernel was initially  calculated, ${\cal K}_m$  is not  M\"{o}bius invariant.  The existence of the scheme where the modified kernel  is   M\"{o}bius invariant (M\"{o}bius scheme) was conjectured in  Ref.~\cite{Fadin:2011we} and then proved  in Ref.~\cite{Fadin:2013hpa}, where the transformation of the kernel from the standard form to  the conformal (M\"{o}bius invariant) form  $ {\cal K}_c$ was found explicitly. 

The eigenvalues $\omega (t)$ of the kernel $ {\cal K}_m$
 calculated in the NLO in \cite{Fadin:2011we} are written  as 
\be
\omega (\nu , n)=- a \left(E_{\nu n}+a\,\epsilon _{\nu n}\right),~~~~~\,\,\,
a=\frac{g^2 N_c}{8\pi^2 }\,,
\label{omeganu}
\ee
where $E_{\nu n}$ is the "energy" in the leading approximation~\cite{Bartels:2008sc}, given by 
\be
E_{\nu n}=-\frac{1}{2}\,\frac{|n|}{\nu ^2+\frac{n^2}{4}}+
\psi (1+i\nu +\frac{|n|}{2})+\psi (1-i\nu +\frac{|n|}{2})
-2\psi (1)\,,\,\,\psi (x)=(\ln \Gamma (x))', 
\ee
and the next-to-leading correction $\epsilon _{\nu n}$ can be written as follows:
\[
\epsilon _{\nu n}=-\frac{1}{4}\Biggl[\psi ^{\prime \prime}\biggl (1+i\nu +\frac{|n|}{2} \biggr )+
\psi ^{\prime \prime} \biggl (1-i\nu +\frac{|n|}{2} \biggr )+\frac{2i\nu \Bigl (\psi ' \bigl (1-i\nu +\frac{|n|}{2} \bigr )-\psi ' \bigl (1+i\nu
+\frac{|n|}{2} \bigr )\Bigr )}{\nu ^2+\frac{n^2}{4}}
\Biggr]
\]
\be
-\zeta (2)\,E_{\nu n}-3\zeta (3)-\frac{1}{4}\,\frac{|n|\,\bigl (\nu
^2-\frac{n^2}{4}\bigr )}{\left(\nu
^2+\frac{n^2}{4}\right)^3}\,.
\label{eigennext}
\ee 
Here the $\zeta(n)$ is the Riemann zeta-function. 

Recently in Ref.~\cite{Bondarenko:2015tba} the  modification of the eigenvalues \eqref{eigennext} was suggested so that they acquired the property of Hermitian separability present for the singlet BFKL kernel \cite{Kotikov:2002ab}.  After this modification the adjoint NLO BFKL eigenvalues are  expressed through holomorphic and antiholomophic parts of the leading order eigenvalue and their derivatives. It was argued that the
proposed choice of the modified NLO expression is supported by the fact that it is
possible to obtain the same result in a relatively straightforward way directly from
the singlet NLO BFKL eigenvalue replacing alternating series by series of constant
sign. 

\section{The modification of the kernel}
The proposed modification of the  eigenvalues \eqref{omeganu}-\eqref{eigennext} is 
\[
\omega(\nu, n)\rightarrow   \omega(\nu, n)+ \Delta\omega(\nu, n)\, , 
\]
\[
\Delta\omega(\nu, n) =\frac{a^2}{2}
\left( \frac{i\nu |n|}{(\nu ^2+\frac{n^2}{4})^2}- \psi ' \biggl(1+i\nu +\frac{|n|}{2} \biggr )+\psi ' \biggl (1-i\nu
+\frac{|n|}{2} \biggr )\right)
\]
\be 
\times \left( \psi (1+i\nu +\frac{|n|}{2})-\psi (1-i\nu +\frac{|n|}{2})\right). \label{addition}
\ee
Evidently, the difference in  the eigenvalues means the difference in  the kernels: 
\[
\hat{\cal K}  \rightarrow  \hat{\cal K} + \Delta \hat{\cal K}~. 
\]
Formally, one can write
\[
\Delta\hat{\cal K}=\sum_{n=-\infty}^{n=+\infty}\,\int_{-\infty}^{+\infty}
{d\nu}\Delta\omega(\nu, n)|\nu, n\rangle\langle \nu, n|~,
\]
where  $|\nu, n\rangle$ are the eigenstates of the kernel  normalized
as
\[
\langle \nu', n'|\nu, n\rangle =\delta_{nn'}\delta(\nu'-\nu)\;.
\]
In M\"{o}bius scheme, the eigenfunctions  $\langle \q_1, \q_2|\nu, n\rangle =\delta(\q-\q_1 -\q_2)\phi _{\nu, n}(\q_1,  \q_2)$ in   the momentum space can be taken as in Refs.~\cite{Fadin:2011we} and \cite{Fadin:2013hpa}, {\it {i.e.}} as    
\be
\phi _{\nu, n}(  q_1,  q_2)=f _{\nu, n}(\frac{q_1}{q_2})=\frac{1}{\sqrt{2\pi^2}}\left(\frac{q_1}{q_2}\right)^{\frac{n}
{2}+i\nu}\left(\frac{q^*_1}{q^*_2}\right)
^{-\frac{n}{2}+i\nu}~,  \;\;\;~~~ q_1+q_2=q\,, \label{conformal eigenfunctions}
\ee
with the normalization
\be
\int \frac{{\qs}d\q_1}{\qs_1\qs_2}\left(\phi _{\nu, n}(  q_1,  q_2)\right)^*\phi _{\mu, m}
(q_1,q_2) =\int \frac{d^2 z}{|z|^2}\, f^{*}_{\nu, n}(z)\, f_{\mu,  m}(z)
=\delta (\mu -\nu )\,\delta_{mn}\,. 
\ee
Here we use the complex notations $q=q_x+iq_y$ and $q^*=q_x-iq_y$.  Then,  we can present the difference in the  kernel as follows:
\be
\Delta K_c(\vec q_1, \vec q_1^{\;\prime}; \vec q)=\sum_{n=-\infty}^{n=+\infty}\,\int_{-\infty}^{+\infty}
{d\nu}\;\Delta \omega(\nu, n)\phi _{\nu n}\,(  q_1,  q_2)\left(\phi _{\nu n}(q'_1,q'_2)\right)^*
\,.\label{Delta K c as sum and integral}
\ee
Let us define 
\be
f_1(z)=\sum_{n=-\infty}^{n=+\infty}\int
{d\nu}\;\frac{1}{2\pi^2}|z^2|^{\,i\nu}\Bigl (\frac{z}{z^*}\Bigr )^{\,\frac{n}{2}} \,\frac{i\nu |n|}{(\nu ^2+\frac{n^2}{4})^2}
\Biggl (\psi \biggl (1+i\nu +\frac{|n|}{2} \biggr )- \psi \biggl (1-i\nu +\frac{|n|}{2} \biggr )\Biggr )~ \label{f1}
\ee
and 
\[
f_2(z)=\sum_{n=-\infty}^{n=+\infty}\int
{d\nu}\;\frac{1}{2\pi^2}|z^2|^{\,i\nu}(\frac{z}{z^*})^{\,\frac{n}{2}} 
\Biggl ( \psi ' \biggl (1-i\nu +\frac{|n|}{2}\biggr )-\psi ' \biggl(1+i\nu+\frac{|n|}{2} \biggr )\Biggr )
\]
\be
\times\Biggl ( \psi \biggl(1+i\nu +\frac{|n|}{2} \biggr)-\psi \biggl (1-i\nu +\frac{|n|}{2} \biggr )\Biggr ),  
\label{f2}
\ee 
so that 
\be
\Delta K_c(\vec q_1, \vec q_1^{\;\prime}; \vec q)= \frac{a^2}{2} F(z), \;\;  F(z) = f_1(z)+f_2(z)~, 
\ee 
where $z=q_1q'_2/(q_2q'_1)$. 

At $|z|< 1$ the integrals over $\nu$ in Eqs.~\eqref{f1} and \eqref{f2} can be calculated by taking residues  in the lower half-plane of  $\nu$. Taking into account that $\psi(x)$ is an analytical function of $x$  having only  poles with residues equal to $-1$ at $x=-l$, $l$ being a natural number,  we obtain for $f_1(z)$ 
\[
f_1(z)=\frac{1}{2\pi}\sum_{n=1}^{\infty}z^n\biggl[\ln|z^2|(\psi (1)-\psi (1 +n)) -\psi'(1 +n)-\psi'(1) 
\]
\be
+\sum_{l=0}^{\infty}|z^2|^{l+1}\left(\frac{1}{(l+1)^2}-\frac{1}{(l+n+1)^2}\right)\biggr]+ c.c.~,
\ee
where {\it {c.c.}} means complex conjugate. Using the relations 
\[
\sum_{n=1}^{\infty}a^n(\psi (1 +n)-\psi (1))=-\frac{\ln(1-a)}{1-a}~, \;\;\;~~
\sum_{n=1}^{\infty}a^n(\psi'(1 +n)+\psi' (1))=\frac{2a\zeta(2)-Li_2(a)}{1-a}~, 
\]
\be
\sum_{l=0}^{\infty}\frac{a^{l+1}}{(l+1)^2}=Li_2(a)~,\;\;\;~~~
 \sum_{n=0}^{\infty}z^n\sum_{l=0}^{\infty}|z^2|^{l+1}\frac{1}{(l+n+1)^2}=z^*\frac{Li_2(z)-Li_2 (|z^2|)}{1-z^*}~, 
\ee
where
\be
Li_2(x)=-\int_{0}^{1}\frac{dy}{y}\ln(1-xy), \;\;~~~~ Li_2(1)=\zeta(2)~, 
\ee
we obtain 
\[
f_1(z) = \frac{1}{2\pi}\left[ \ln|z^2|\left(\frac{\ln(1-z)}{1-z}+\frac{\ln(1-z^*)}{1-z^*}\right)+2\frac{1-|z^2|}{|1-z|^2}Li_2(|z^2|)\right. 
\]
\be
\left.+\frac{1-2z^*+|z^2|}{|1-z|^2}Li_2(z)+\frac{1-2z+|z^2|}{|1-z|^2}Li_2(z^*) -2\frac{z+z^*-2|z^2|}{|1-z|^2}\zeta(2)\right]. 
\ee
For the function $f_2(z)$, taking into account that 
\be
\psi(x)\psi'(x)|_{x\rightarrow -l} =-\frac{1}{(x+l)^3}+\frac{\psi(1+l)}{(x+l)^2} +constant~,  
\ee
where $l$ is a natural number, we have  at $|z|< 1$ 
\[
f_2(z)=\frac{1}{2\pi}\sum_{n=0}^{\infty}\left(1-\frac12\delta_{n,0}\right)z^n
\]
\be
\times \sum_{l=0}^{\infty}|z^2|^{l+1} \left[\ln^2|z^2| +2\ln|z^2|\left(\psi (1 +l)-\psi(2+n+l)\right)-4\psi'(2+n+l)\right]+ c.c.~.
\ee  
The  equalities   
\[
\sum_{l=0}^{\infty}a^{1+l}\psi(2+l) = \frac{a\psi(1)-\ln(1-a)}{1-a}~,\;\;\;~~~
\sum_{l=0}^{\infty}a^{1+l}\psi'(2+l) = \frac{a\zeta(2)-Li_2(a)}{1-a}~,\;\;\;
\]
\[
\sum_{n=1}^{\infty}z^n\sum_{l=0}^{\infty}|z^2|^{l+1}\psi(2+n+l)=
\frac{1}{1-z^*}\left[\frac{\ln(1-|z^2|)-|z^4|\psi(1)}{1-|z^2|} -z^*\frac{\ln(1-z)-z^2\psi(1)}{1-z} \right]~,\;\; 
\]
\be
\sum_{n=1}^{\infty}z^n\sum_{l=0}^{\infty}|z^2|^{l+1}\psi'(2+n+l)
=
\frac{1}{1-z^*}\left[\frac{Li_2(|z^2|)-|z^2|\zeta_2}{1-|z^2|} -z^*\frac{Li_2(z)=z\zeta_2}{1-z^2} \right]~\;\; 
\ee
give us 
\[
f_2(z) = \frac{1}{2\pi|1-z|^2}\Bigl[ |z^2|\ln^2|z^2| -2\ln|z^2|\Bigl((1+|z^2|)\ln(1-|z^2|)
\]
\be
-z^*\ln(1-z) -z\ln(1-z^*)\Bigr) -4Li_2(|z^2|)+4z^*Li_2(z)+4z Li_2(z^*) -4|z^2|\zeta(2)\Bigr]~. 
\ee
For the sum $F(z)=f_1(z)+f_1(z)$ we obtain  at $|z|< 1$ 
\[
F(z) = \frac{1}{2\pi(|1-z|^2)}\Bigl[ |z^2|\ln^2|z^2| -\ln|z^2|\Bigl(2(1+|z^2|)\ln(1-|z^2|) 
\]
\[
-(1+z^*)\ln(1-z) -(1+z)\ln(1-z^*)\Bigr) -2(1+|z^2|)Li_2(|z^2|)
\]
\be
+(1+2z^*+|z^2|)Li_2(z)+(1+2z+|z^2|) Li_2(z^*) -2(z^*+z)\zeta(2)\Bigr]~.  \label{F(z)}
\ee 

Here it should be noted that  the functions $f_1(z)$ and  $f_2(z)$ (and hence their sum $F(z)$) are defined by Eqs.~\eqref{f1} and  \eqref{f2}  both for $|z|<1$ and for $|z|>1$; moreover, due to the property $\Delta\omega(\nu, n)$ (see Eq.~\eqref{addition}) 
\be
\Delta\omega(-\nu, -n) = \Delta\omega(\nu, n)~, 
\ee
it must be 
\be
F(z) = F\left(\frac{1}{z}\right)~.  \label{reverce} 
\ee
Together with  Eq.~\eqref{F(z)}, which  gives  the function $F(z)$ in  the region $|z|<1$,  Eq.~\eqref{reverce}  determines $F(z)$ in the region   $|z|>1$.  From the other hand, the right side of Eq.~\eqref{F(z)} gives the function ${\cal F}(z)$  in the whole plane of $z$.  It turns out, however, that at $|z|>1$ the function  
${\cal F}(z)$  does not coincide with  $F(z)$ determined  by  Eq.~\eqref{reverce}. Indeed, it is seen from Eq.~\eqref{F(z)} that the function ${\cal F}(z)$ has a cut starting at $z=1$ and is not a single valued function. To see this clearly one can rewrite ${\cal F}(z)$,  using the relation 
\be
Li_2(x)+Li_2(1-x)=\zeta(2)-\ln(x)\ln(1-x)~,
\ee
in the form  
\[
{\cal F}(z) = \frac{1}{2\pi(|1-z|^2)}\biggl[ |z^2|\ln^2|z^2|+2(1+|z^2|)Li_2(1-|z^2|)-(1+2z^*+|z^2|)Li_2(1-z) 
\]
\[
-(1+2z+|z^2|) Li_2(1-z^*)+\frac12(1-|z^2|)\ln|z^2|\ln|1-z|^2 +\frac{z-z^*}{2}\ln\frac{z}{z^*}\ln|1-z|^2 
\]
\be
-\frac{|1+z|^2}{2}\ln\frac{z}{z^*}\ln\frac{1-z}{1-z^*}\biggr]~. \label{F1(z)}
\ee
It is easy to see that all terms besides the last one are single valued around the point $z=1$, but the last one has not such property. Of course, ${\cal F}(z)$ is a single valued function at $|z|<1$; but this property is lost in the whole $z$-plane. 
It means, in particular,  that ${\cal F}(z)\neq {\cal F}(\frac{1}{z})$. This can be shown explicitly 
from Eq.~\eqref{F1(z)} using the relation 
\be
Li_2(1-x)+Li_2(1-1/x)=-\frac12 \ln^2(x)~. 
\ee
It gives 
\[
{\cal F}\left(\frac{1}{z}\right) = \frac{1}{2\pi|1-z|^2}\biggl[-\frac{|1-z|^2}{4}\ln^2|z^2| -2(1+|z^2|)Li_2(1-|z^2|)
\]
\[
+(1+2z+|z^2|)Li_2(1-z) +(1+2z^*+|z^2|) Li_2(1-z^*)+\frac12(1-|z^2|)\ln|z^2|\ln|1-z|^2 
\]
\be
+\frac{z-z^*}{2}\ln\frac{z}{z^*}\ln|1-z|^2 -\frac{|1+z|^2}{4}\ln^2\frac{z}{z^*}+\frac{|1+z|^2}{2}\ln\frac{z}{z^*}\ln\frac{z-1}{z^*-1}\biggr]~. \label{F(1/z)}
\ee
Note that the point $z=1$ is the only singular  point of the function ${\cal F}(z)$  in the closed circle  $|z|\le 1$. Moreover, it is easily  seen from Eq.~\eqref{F1(z)} that the singularity of ${\cal F}(z)$ in this  point  is an integrable one.  It means that the modification of the eigenvalues \eqref{addition} is related
only with the real part of the kernel.  Thus, we obtain that the modification of the BFKL kernel corresponding to the modification of the eigenvalues suggested in Ref.~\cite{Bondarenko:2015tba}  is written as  
\be
\Delta K_c(\vec q_1, \vec q_1^{\;\prime}; \vec q)=
 \Biggl\{\begin{array}{ll}\frac{a^2}{2}F\left(\frac{q_1q'_2}{q_2q'_1}\right) & \mbox{if $\Big|\frac{q_1q'_2}{q_2q'_1}\Big| \le 1$}\\
\frac{a^2}{2}F\left(\frac{q_2q'_1}{q_1q'_2}\right) & \mbox{if $\Big|\frac{q_1q'_2}{q_2q'_1}\Big| \ge 1$ } 
\end{array}, \label{correction}
\ee
where $F(z)$ is defined  in  Eq.~\eqref{F(z)}, $q_1+ q_2= q'_1+q'_2 =q$, 
and  can not be presented by a single analytic function in the entire domain.

\section{Conclusion}
We found the correction \eqref{correction} to the BFKL kernel for the  adjoint representation of the colour group in the planar N=4 SYM corresponding to the modification of  the eigenvalues of the kernel suggested in Ref.~\cite{Bondarenko:2015tba}.  It turned out that this  correction  is related only  to the real part of the kernel. However, it  can not be presented by one analytic function in the entire region of transverse momenta, contrary to  the real parts of the  kernel in the M\"{o}bius \cite{Fadin:2013hpa}  and  standard \cite{Fadin:2000kx}-\cite{Gerasimov:2010zzb} schemes. Note that the real part in the standard scheme was found for arbitrary space-time dimension, 
therefore the argument of Ref.~\cite{Bondarenko:2015tba}  in favour of the modification, based on removal of the infrared divergences  seems untenable.  

In our opinion, other arguments of  Ref.~\cite{Bondarenko:2015tba} in favour of the modification are also inconsistent. The ambiguity of the NLO kernel because of the possibility to  redistribute  NLO corrections between the kernel and impact factors is irrelevant, because transformations of the kernel admitting to this ambiguity do not change their eigenvalues. It is clearly  seen from the fact that change  of eigenvalues means change of dependence on energy, whereas impact factors are energy independent by definition. 
It was argued  also  in Ref.~\cite{Bondarenko:2015tba} that   the modification is  supported by the fact that it is possible to obtain the same result in a relatively straightforward way directly from the singlet NLO BFKL eigenvalue, replacing alternating series by series of constant
sign. But it can not be a serious argument because there is no simple relation between singlet and adjoint kernels. 

Thus,  the modification of the eigenvalues  of the BFKL kernel suggested in Ref.~\cite{Bondarenko:2015tba} contradicts to the known properties of the kernel, and the main motivation for this modification --  the Hermitian separability of the  eigenvalues -- does not have serious grounds. 

\vspace{0.5cm} {\textbf{{\Large Acknowledgments}}}

\vspace{0.5cm} V.S.F. is grateful to L.N.~Lipatov for discussions and  thanks the Dipartimento di Fisica
dell'Universit\`{a} della Calabria and the Istituto Nazionale di
Fisica Nucleare (INFN), Gruppo Collegato di Cosenza, for the warm hospitality
while part of this work was done and for financial support.

\end{document}